\documentclass[twocolumn,showpacs,preprintnumbers]{revtex4-1}
\usepackage{graphicx}
\usepackage{dcolumn}
\usepackage{bm}
\usepackage{color}
\usepackage{amsmath}
\usepackage{amsfonts}
\usepackage[pdftex]{epsfig}
\usepackage{epstopdf}
\usepackage{float}
\usepackage{subfigure}
\usepackage{caption}

\begin{document}

\title{Pb induces superconductivity in Bi$_2$Se$_3$ analyzed by point contact spectroscopy}

\author{P. Arevalo-L\'{o}pez}
\author{R. E. L\'{o}pez-Romero}
\author{R. Escudero(1)}
\affiliation{Instituto de Investigaciones en Materiales, Universidad Nacional Aut\'{o}noma de M\'{e}xico. A. Postal 70-360. M\'{e}xico, D.F., 04510 M\'EXICO.}

\date{\today}

\begin{abstract}
Some topological insulators  become superconducting when doped with Cu and Pd. Superconductivity in a  non-superconductor   may be induced by   proximity effect: i.e.  Contacting    a non-superconductor  with  a superconductor.  The superconducting  macroscopic wave function will  induce electronic pairing  into the normal compound. In the   simplest  topological insulator, Bi$_2$Se$_3$,  superconductivity may be  induced with  Pb. We  studied with point contact  junctions  formed by contacting  Bi$_2$Se$_3$  crystals and Pb,  glued and pressed with  silver paste and/or applying low heat to the Pb to improve the contact.  Junctions were formed with   a  thin tungsten plated gold W(Au) wire as one  electrode, and the other $Bi_2Se_3$ and Pb.  We study the characteristics of the electron coupling; the transition temperature,$T_C$,  evolution with temperature of the energy gap, $\Delta$,  and  2$\Delta$/K$_B$T$_C$  ratio. The  superconductor  Bi$_2$Se$_3$-Pb behaves  different as explained in the classical  BCS theory. In BCS  a superconductor is only   weak or strong coupled, depending on the electronic interaction  that form the  electronic condensate. This  differentiation is given by the size of the  mentioned  ratio.  BCS Typical values are  3.53 to about 4.3 for weak or strong coupling limits respectively.  In this study  performed in Bi$_2$Se$_3$-Pb we found  different values   to the normal ones  from  10  to 23,  indicating  very strong limit. Those  values    never have  been observed in other  superconductor. The  transition temperatures  found varies  from 2.7  to  7 K. This information and  other results will be presented in this paper.
\end{abstract}

\maketitle

\section{Introduction}
With the discovery of topological insulators (TI) \cite{xia,chen2,hasan} many theoretical and experimental studies have been made  in order to  improve our knowledge  of its  new properties. In this work we are interested in  the superconducting behavior induced by the so called proximity effect studied with point contact junctions (PCS),  and in the new properties of these  TI superconductors \cite{Qui, wrat2010, kriener, Hor2010, Hor, hsieh, sat}. In the new TI; Bi$_2$Se$_3$ and Bi$_2$Te$_3$  was found that both compounds become superconducting when copper or palladium are doped or intercalated  in  the structural gaps between TI layers  \cite{ kriener, Hor2010, Hor, sat, par, koren, wang,  Desheng, tet}. In these superconductors  the transition temperatures are low, $T_C \sim 3 - 5.5$ K \cite{kriener, Hor2010, Hor, sat}. Actually many work with TI are been   performed  to determine other metals that can produce superconductivity. The interest  in these new superconductors is  to  found  the general behavior and to  understand if they  present similar electronic pairing   mechanism as given by  the BCS theory  \cite{Hor,hsieh,sat}. In BCS  the  electron pairing is due to   the  interaction between two electrons at the Fermi surface by a  phonon exchanged  of the lattice  structure.  Among the important parameters and variables of the  theory, one is very   related to the pair formation,  the 2$\Delta$/K$_B$T$_C$ ratio, where $\Delta$ is the energy gap, $T_C$ is the transition temperature, and $K_B$  the Boltzmann constant. In  BCS theory  two possibilities exist for the  strength of the  pair formation,  the  weak or strong coupling limit.  The  weak limit is given by the number 3.53, whereas  for  the strong  limit the value is  slightly  above the  weak.  For instance  Pb,  a typical strong coupled superconductor the coupling is 4.3. Many more values for  the ratio can be  found in Table 5.2 in Poole book for conventional superconductors \cite{Poole}.  In  unconventional superconductor  the behavior may be  very different, but no so,   as Cuprates, Arsenides, Fe  based, etc.  In these compounds the ratio varies from  7 to about 8, although some of those  values may be controversial as estipulate in different publications. However this  ratio has  ever found  bigger than 10 \cite{daghero, schesinger, dubroka, manzke, lee, escudero1}. The conventional wisdom related to big  values of  the mentioned ratio is that superconductors with different type of electron pairing  could  have  different BCS microscopic mechanism.

\begin{figure}
\centerline{\includegraphics[scale=0.35]{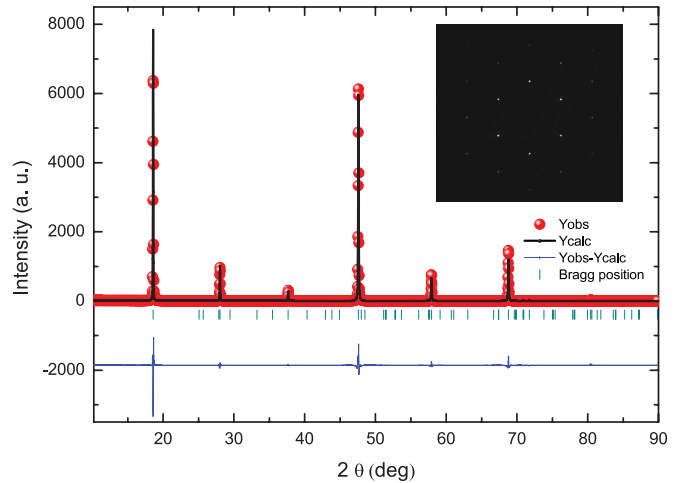}}
\caption{(Color online) X- ray diffraction data of the Bi$_2$Se$_3$ single crystal, and Rietveld refinement. Only the (00$l$) reflections  are observed. Inset shows  electron diffraction data of a  thin layer of the   crystal. }
\label{fig1}
\end{figure}

One of the main reasons to study superconductivity in TI induced by  proximity effect in Bi$_2$Se$_3$ is that it is the simplest one found at present, with  simple Dirac cone  and large band gap, it has a very simple  surface states, which is quite different to  Bi$_2$Te$_3$, the other well known TI.  this last TI  has a large deviation of the Dirac cone and this  deviation can  introduce extra   difficulties to understand the induced superconductivity.  Hasan has pointed out  that  in this TI may be induced  a two dimensional superconductor state  \cite{hasan},  in similar form as occurs in the high $T_C$ superconductors.  i.e., this implies that the majorana states may be manipulated \cite{Fu} and  including the possibility to observe features and effects related to  those modes, and as final result different superconducting mechanisms  related  to  different electronic coupling. This assumption was discussed by Qi and Zhang supposing that in this TI could  be induced a non trivial topological superconductor and perhaps to be one of the ways to probe the existence of  majorana modes \cite{Qui}.

For the study of the induced superconductivity in Bi$_2$Se$_3$ we  choose PCS junctions as  the experimental tool. Tunneling and point contact spectroscopies are simple tools and well appropriate  to  study   superconductivity. The compound was prepared with Bi$_2$Se$_3$ contacted with Pb. A Pb foil  induce the  proximity effect because the macroscopic wave function of the superconducting state produces the pair formation inside the TI when  two materials  are  in very close proximity. Our study was performed  using point  contact junctions (PCS). These junctions could  give information about the microscopic mechanisms for the pair formation, and  features and structure of the Andreev reflections,  and information about the energy gap.

  Before to start the study,  it is important  to determine the working regime of our PCS junctions. From  theoretical bases we determined  the  junctions working regime  using the BTK, model,  which is  well known  \cite{BTK}. In this model  the microscopic parameter used to   determine and  characterize  the junctions parameters can indicates if it behaves as  tunnel or metallic contact. $Z$ is the parameter used which  according to the value, determines the junctions behavior. A small  value,  close to  zero means junctions with  pure Andreev behavior thus,  a metallic contact. Whereas values above zero mark the transition from metal contact to tunneling regime. Value from above 1.5 may be considered with  almost tunneling behavior.

Preparation of the single crystal Bi$_2$Se$_3$ were produced by mixing stoichiometric amounts of high purity Bi pieces, (Sigma-Aldrich $99.999\%$), and Se (Alfa Aesar $99.999\%$),   were grounded, pressed into pellets and sealed in evacuated quartz tubes, heated for about 9 $h$ to a maximum temperature of 850 $^{\circ}$C and held at this  temperature for about 24 $h$. After this procedure the tube was slowly cooled to 630 $^{\circ}$C and annealed for 24 $h$. The last step was performed by quenching the tube into water at room temperature. The  single crystals present a plaquette-like shape,  length sizes about  0.5 - 1 cm, and 0.1 to 0.5 width. Phase identification was done with  X-ray diffractometer, Bruker D8  in a Bragg - Brentano geometry. Also  electron diffraction microscopy  was performed. For this study  mechanical exfoliation was performed.

Surface morphology was determined in a scanning electron microscope FEG JEOL-7600F equipped with X-ray energy dispersive analyzer Oxford INCA X-Act. Selected-area of electron diffraction patterns were obtained in a Transmission Electron Microscope JEOL JEM-1200EX.
The crystal structure was confirmed by X-ray diffraction, Fig. 1,  by comparison  with X-ray of Bi$_2$Se$_3$ and Database (ICSD: $04-2545$). The patterns show a single phases. Reflections are very sharp indicating the high crystallinity and (00$l$) reflections from the basal plane of  cleaved sample and absences for the $R\bar{3}m$ rhombohedral space group.

    The structural characteristics  of   Bi$_2$Se$_3$ are the following; it  has a rhombohedral crystalline structure with space group $R\bar{3}m$ \cite{Zhang2009,Hsieh2009}, and   parameters as  $a$ = 4.143 \AA,  and $c$ = 28.636 \AA. The structure is formed by layers, with  three double-layers of sharing-edges Bi$_2$Se$_6$ octahedra  stacked along the [00$l$] direction, equivalent to the $Se1-Bi-Se2-Bi-Se1$ quintuple layers as often referred in the literature \cite{ Hor, Wiese1960,Zhang2009b}. These double-layers are weakly bonded via van der Waals interactions, resulting in an easy cleavage material \cite{Wiese1960}. According, the structural characteristics Bi$_2$Se$_3$ may accept many chemical modifications either by substitution  or by intercalation between the double-layers.
    To induce  superconductivity by proximity effect we used   a thin  Pb foil,  pressed and/or  glued to the TI with silver paint or by mechanical attachment with  low heating as well.  Similar experiments have been performed using mechanical gluing  with different high or low transition temperature  superconductors \cite{par,koren,Duming,Fan,Koren2013}. The spectroscopic characteristics   determined with PCS junctions, were  formed with  Bi$_2$Se$_3$-Pb (TI-Pb) as the main electrode, and  the other one  to complete the junction was a  thin wire of W(Au) 5 $\mu$-m diameter. The spectroscopic characteristics,   the Andreev reflections, transition temperatures, energy gaps, etc,  were observed at low temperature. The number of studied  PCS's  shown  different transition temperatures, depending in  the close contact  in both materials. The energy gap extracted with  the measured differential conductance characteristic, $dI/dV$, showed the  evolution  of the energy gap with temperature. The transition temperature,  evolution with temperature of the energy gap,  and the $2\Delta/K_BT_C$  ratio were determined. The ratio gives  values quite anomalous and big showing values  from  $10$ to  $23$. Additionally, extra  information was obtained by determining the spectral features of the structure in the derivative of the differential resistance obtained at the lowest temperature. Clearly the  structure may be  related to the  phonon  spectra of the two components, the Bi$_2$Se$_3$ and  the Pb. The correlation of effects related to phonon structure were observed and determined in PCS by Yanson  many years ago \cite{yanson}.

\begin{figure}
\includegraphics[scale=0.3]{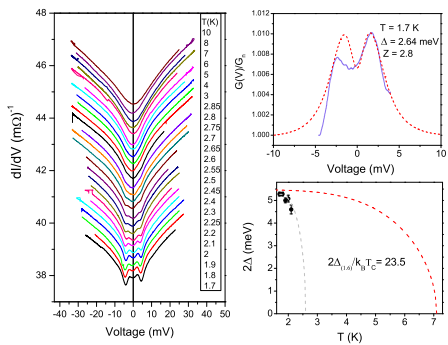}
\caption{(Color online) Differential  conductance of a Bi$_2$Se$_3$-Pb junction  from 1.7 to 10 K, left panel. At low temperature  are clearly noted the  spectral characteristics of the energy gap feature. Above 3 K small features are  distinguished  related to the superconducting Pb. Top right  panel shows  BTK fit of $G(V)/G_n$ - $Bias$ $voltage$   of the energy gap features at  $T$= 1.7 K  induced by Pb, parameters were;  $\Delta$ = 2.64 meV, $Z$ = 2.8. The right lower panel shows the evolution of the energy gap with temperature. The red red line) is BCS for Pb,  $2\Delta/K_BT_C$ ratio is this junction was 23.5. }
\label{fig2A}
\end{figure}

\begin{figure}
\includegraphics[scale=0.3]{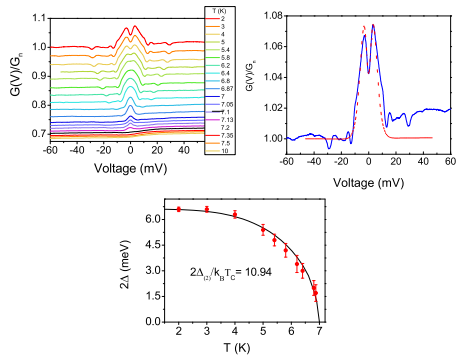}
\caption{(Color online) Left panel shows the differential conductance of a  PCS junction, the  transition temperature was  very  close to the Pb transition. All  curves were  vertically displaced by a small amount  in order to  have a clear   view of the spectroscopic features.  Right panel shows the fit with BTK model values were  $\Delta$ = 3.3 meV,  and $Z$ = 0.73 at 2 K. The lower panel shows  the temperature evolution of the energy gap. The lower panel also shows the anomalous big  $2\Delta/K_BT_C$ ratio .}
\label{fig3A}
\end{figure}

\section{Results and Discussion}

 The characteristics of the TI cell parameters were determined by Le Bail fitting,   $a$ = 4.1265(1) \AA and $c$ = 28.6349(2) \AA. Results related to the structural characterization are  in Fig. 1. There we present the X-ray  and   Rietveld refining.  The inset is the spectrum of electron diffraction of a thin layer of the single crystal used in this experimental work. Data and  spectrum show  good structural characteristics of the single crystal.

 All  PCS's junctions studied  were formed by a  small crystal layers of Bi$_2$Se$_3$, the thickness  about  $0.1 - 0.35$ mm, and a foil  of Pb with similar size of the  Bi$_2$Se$_3$. The two materials, Pb and TI were pressed together and glued with  diluted silver  paint, and sometime a low heat was applied to Pb in order to increase the contact. Both materials were pressed with a small press  to increases the contact.  This simple manner was used to   maintain together  the  two materials  and  induces the proximity  effect in the Bi$_2$Se$_3$. As mentioned PCS's were fabricated by using a thin wire of W(Au), 5 $\mu$m diameter contacting the  TI-Pb sample  on the TI surface. Junctions differential  resistances varied from  1 to 25 $\Omega$.  Other PCS's  with very small differential resistances, in the order to   1 to 5  $\Omega$ were discarded because heating effect distorted the spectral characteristics and were unstable,  only high resistance values, about  20 - 25 $\Omega$ were used avoiding heating effects. The number of PCS's fabricated were sufficient to obtain a clear picture and reproducibility. Measured PCS's were more than  30 - 40. Only  well reproducible results were taken into account. Differential resistance ($dV/dI - V$) was obtained with the normal tools  for  tunneling studies; Lock-in amplifier and bridge,  the inverse of ($dV/dI - V$),  $G(V)$, was digitally obtained. As is well known  $G(V)$ at low temperatures  is proportional to the density of electronic states, and information about the gap was  obtained.

 Figures 2 to 4 show  result  of  our measurements performed in   three different PCS's. The  differential conductance of these PCS's have  similar  values,  but presents   different spectroscopic features, which will be related  to the coupling of  the TI-Pb system. The spectroscopic features of the PCS  varies  because the closeness contact in  both  materials. The features are therefore due by the  coupling of the superconducting macroscopic wave function into the normal material. This coupling readily may be  determined by the transition temperature,  size of the energy gap and  $2\Delta/K_BT_C$  ratio,  and  by the  spectroscopic features observed on the background of $G(V-eV)$ from low energies up to relative high values  (similar to occurs with the determination of phonon spectrum in normal superconductors) and low temperature,    after the subtraction  of the energy gap was performed.
 Figure 2 is  the result of a measured PCS with  $Z = 2.8$  and $\Delta = 2.64$ meV values at a temperature of 1.7 K.  Right  panel   shows the fitting of  the normalized differential conductance ($G_n$ determined  at 1.7 K.  In this figure the data is the  continuous line   whereas  the pointed line  is the fit to the  BTK model.

  Fig. 2 right lower panel displays  the temperature evolution of the  energy gap, two curves are shown; the curve that displays the  energy gap at about 2.5 K is the result of the proximity effect in TI-Pb system. We displayed  the BCS evolution of the energy gap with temperature of the Pb, for reference  to  emphasize the similar values of the energy gap in this TI-Pb system. The $2\Delta/K_BT_C$  ratio is incredible big, about 23.5. This value  was the biggest in all our experiments.

  Fig. 3 shows another PCS's data measured from 2 to 10 K, left panel.  The energy gap features are clearly observed at low temperatures and disappears at about 7.1 - 7.2 K,  Above 10 K no structural features are observed.   As in the other junctions characteristics at low temperature the  gap feature, clearly is seen  at  the center of the curves,  close to  zero bias, and  decreasing with the increasing temperature. This figure also shows the  normalized data measured at 2 K,   Panel right shows BTK fitting with $Z= 0.73$, and $\Delta$ = 3.3 meV. This PCS  behaves as a metallic contact, with  spectroscopic features only related to  Andreev reflections \cite{yanson}.  More information was obtained from data of the same PCS. Above the energy gap bias voltage the  structure in the Differential conductance clearly may be   related  to phonon spectrum of the TI-Pb system, as explained many years ago by Yanson when studying in point contact junctions  \cite{yanson}.
   It is interesting to pointing out another characteristics seen  in  our  junctions. Close to the zero bias voltage the features related to the energy gap displayed a  non symmetric structure. This can be see in the Data of Fig. 3, in the   right panel  the two peaks have different values, one bigger to  the another. These nonsymmetric peaks are shown in figures 2 and 3. The  two  conductance peak of the energy gap feature  show different  conductance values. This  anomalous characteristic looks quite  similar to the  observed and found in heavy  superconductor and in other physical  systems. For   example  similar characteristics  were found in URu$_2$Si$_2$ \cite{escudero}. The  anomaly  was related to a Fano resonance \cite{fano}, and here in these experiments the anomaly  could be explained as an  extra interaction still not considered in the TI.

  Lastly in this figure, the lower panel displays the evolution of the energy gap feature with  temperature. The transition  temperature was 7.1  K, only  a little below  Pb transition. However,  the $2\Delta/K_BT_C$ ratio, as in previous  experiments it  is big,  10.94. Here  at this point we need to analyze the values and size of the ratios: With  lower transition temperatures ratio values are  bigger with small transition temperatures, whereas high transition temperatures shows lower ratios.

\begin{figure}
\includegraphics[scale=0.3]{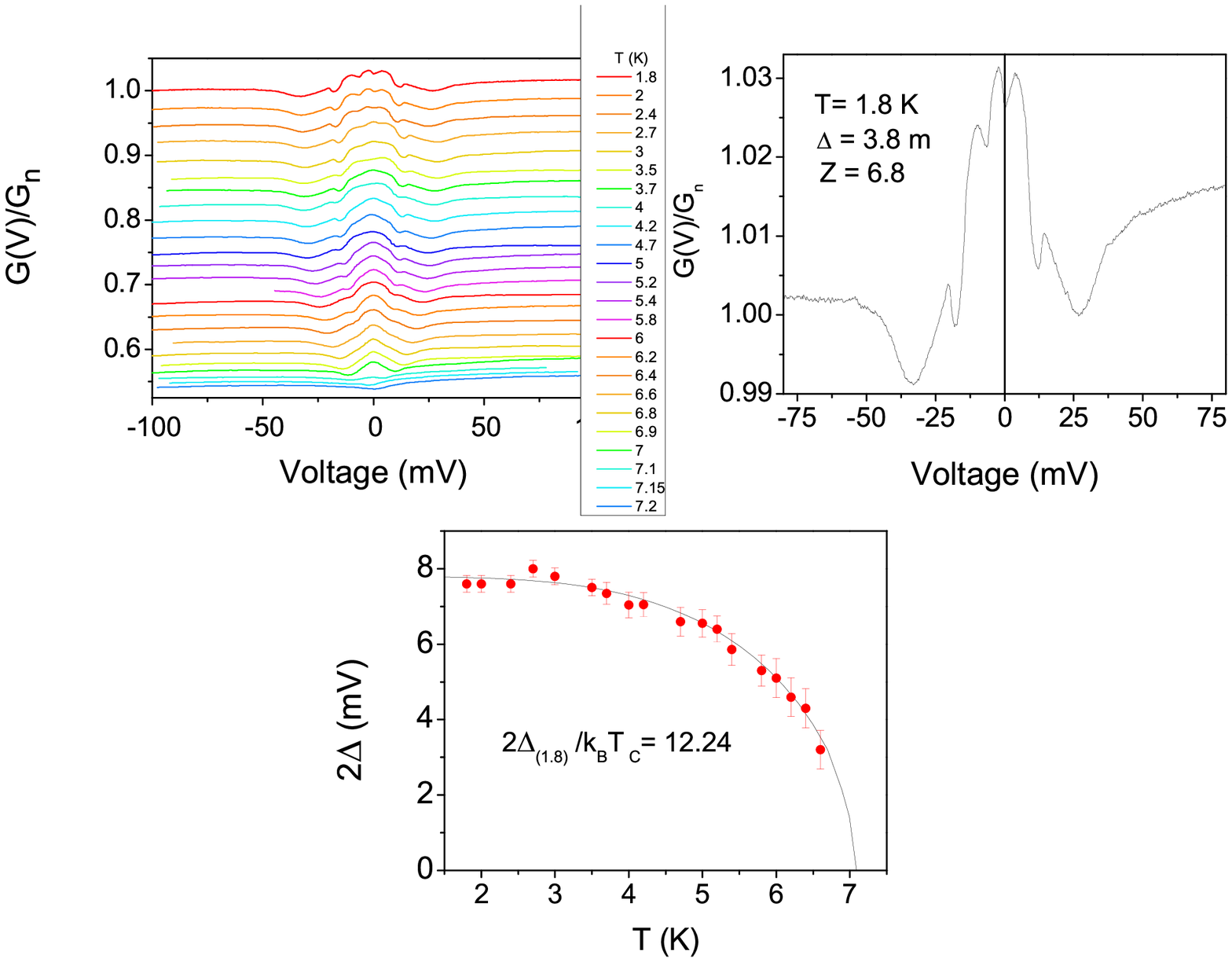}
\caption{(Color online) Normalized differential conductance measured from 1.8 to 7.2 K, left panel and  right panel shows the spectroscopic features with BTK at 1.8 K, note the additional  structure above the energy gap features. Determination of the gap feature gave  $\Delta$ = 3.8 meV, and $Z$ = 6.8,  the black continuous line is experimental data. The lower panel  displays the evolution with temperature of the gap, extracted of the characteristics. The fitting was performed only from low temperature to 6.8 K. However, extrapolation at high temperature indicates features close to 7.2 K. $2\Delta/K_BT_C$ ratio value is   12.24.}
\label{fig4A}
\end{figure}

 The last PCS junctions measured are presented in Figs. 4, it  shows the overall evolution from 1.8 K to  7.2 K.  Close to zero bias voltage are  the features of the energy gap. More structure was   observed here. Differential conductance curve  fitted with BTK at 1.8 K,  gave  $Z = 6.8$ which indicate a tunnel junction,  with gap,  $\Delta$ = 3.8 meV. Again the spectroscopic features of the gap, as in the other case are  anomalous, both  peaks are different in conductance value. However the energy gap evolution as in other  PCS's  follows BCS, Here the detrmined ratio was  12.24.

   More characteristics can be extracted in PCS which include the possibility to observe the phonon spectra. In  this  study we are in the possibility to correlate structure observed with  phonon spectrum in the structure obtained in our PCS's. Fig. 11 shows the spectra "phonon density of states" obtained in four PCS's once that the second derivative of the voltage respect to the current was calculated, $d^2V/dI^2-Bias$ $voltage$-$\Delta$, (this is show in lower inset). The upper panels, we  plotted the normalized differential conductance of Pb obtained by tunneling as in Parks book \cite{Parks}, and the phonon density of states of the Bi$_2$Se$_3$  as given by Rauh \cite{Rauh}. The lower inset displays our results, we noted some similar structure with the real data of the upper inset. There is a   correlation between the two spectrum, the obtained here, and the Pb and Bi$_2$Se$_3$ already known. We believe that further analysis of these spectrum are necessaries  in order to have a better understanding of the manner the two systems are interacting.

\begin{figure}
\includegraphics[scale=0.3]{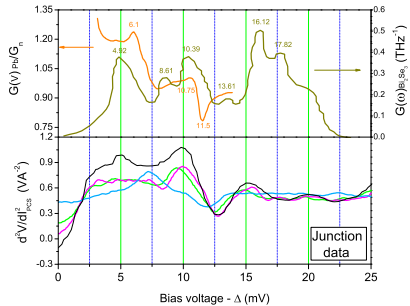}
\caption{(Color online) This figure displays  the density of states of the phonon structure of Pb  in the  upper inset  ( obtained by tunneling measurements (orange line) \cite{Parks}, and the phonon structure of Bi$_2$Se$_3$ as taken of  literature (green line) \cite{Rauh}. Lower inset shows our experimental data, the  second derivative of $dV/dI$  of four junctions with the TI-Pb sample.  Qualitatively it seems a close  concordance with our data. }
\label{fig5}
\end{figure}

Lastly, in this study we found  different  transition temperatures  with the   different junctions studied, the different transition temperatures were because the proximity coupling  due to the  contact closeness of the two materials. The variations in the $T_C$  were from  2.7  to 7.1 K. The coupling  values for  the ratio,   $2\Delta/K_BT_C$,  in  distinct PCS,   were bigger than for Pb, which is a strong coupled superconductor with the bigger values of all  known superconductors.  As  mentioned  the energy gap characteristics were fitted with  the BTK model and  parameter,  $Z$. This varies  from  $\sim$ 0.7 to 7.0.  The results  show  the energy gap evolution with temperature, and follows  BCS theory. In addition, extra  information was obtained with the study of  the second derivative of the voltage  respect to the current ($d^2V/dI^2-Bias$ $voltage$-$\Delta$), which  provides  information related to phonon spectrum. In old studies  with normal superconductors is a generalized test of energy of the phonons that couples the superconducting pairs.     The background  in the early studies performed in tunnel junctions, $G(V) - Bias$ $voltage$ was  related to  the phonon density  interacting to form the superconducting state.  In PCS the initial studies related to phonon spectrum was given by I.K. Yanson  \cite{yanson}.

The results and characteristics of this new Bi$_2$Se$_3$-Pb compound is  resumed in our obtained data. Very strong coupled limits by  proximity effect induced with  Pb were observed. At the moment many questions remain, but it is convenient  to mention that because in  TI the surface sates are symmetry protected, and thus exist particle number conservation and time reversal symmetry, and in addition  in 3D the Fermi surface may be in the conduction or valence band because  the number of defect in natural material. However this $Bi_2Se_3$ it was suggested  that the  surface states may be considered as a new type of 2 dimensional electron gas, therefore  electronic characteristic for the coupling may be similar to the cuprate  superconductors. According this 2-d system   has  to be relatd to the so strong coupling limit \cite{hasan,Qui,Fu}.

In Conclusion;  we studied with PCS's the  system formed by a thin layers of the Bi$_2$Se$_3$ single crystal glued to a foil of Pb and pressed to the TI with diluted silver paint. The  junctions were formed with  Bi$_2$Se$_3$-Pb system, the junction was formed with an W(Au) wire as second electrode  firmly  put on  the surface of the TI. Measurements were performed with the normal tools for tunneling or metallic contacts; lock-in amplifier and bridge,  an MPMS by  Quantum Design used as the cryostat to control the temperature. The characteristics of the junctions were determined at low temperature,  1.7 or 2 K and to  10 K.  In all junctions  the evolution with temperature of the  gap feature follows BCS, with different  transition temperatures.  The junctions characteristics indicate  superconducting proximity effect  with different couplings, depending on the closeness with  Pb. The energy gap value was determined  with  BTK model, and with BCS  parameter,  2$\Delta/K_BT_C$ , this varies  from  10 to 23.  The extreme values of the ratio is  indicative of an anomalous  strong coupling, never observed before. It is important to mention that the anomalous big value of the ratio was observed in all  junctions,  those values indicate the influence of the coupling strength but not in the  transition temperature.  This value of the ratio is not clear in the relationship  of the electronic coupling strength with the transition temperature; The junction with small $T_C$ has the biggest ratio, whereas for similar $T_C$ values the ratio is approximately  equal.  As mentioned by  other researches in  Cu-Bi$_2$Se$_3$  similar big  values were found and without a clear  interpretation of the results \cite{kirz,chen,niv,Yamakage2012}.  We conclude that  in this topic  more theoretical and experimental work are  necessary  to   understand the   new behavior, find correlations in different TI and perform studies with weak and strong coupled superconductors,  using  PCS junctions with simple  superconducting elements as Al, Sn, In.

\begin{acknowledgments}
We thanks to Prof. R. Cava for  kindly provide us the method  to growth the single crystals. For  grant by DGAPA-UNAM, project IN106014. P. Arevalo-l\'{o}pez  thanks to CONACYT for scholarship number 216073,  And also to J. Morales, A. l\'{o}pez.
\end{acknowledgments}


\begin{thebibliography}{99}

\bibitem{xia}Y. Xia, D. Qian, D. Hsieh, L. Wray, A. Pal, H. Lin, A. Bansi, D. Grauer, Y. S. Hor, R. J. Cava and M. Z. Hasan,  Nature Phys. Nature 5, 398 (2009).
\bibitem{chen2}
Y. L. Chen, J. G. Analytis, J.-H. Chu, Z. K. Liu, S.-K. Mo, X. L. Qi, H. J. Zhang, D. H. Lu, X. Dai, Z. Fang, S. C. Zhang, I. R. Fisher, Z. Hussain and Z. X. Shen, Science 325, 178 (2009).
\bibitem{hasan}
M. Z. Hasan and C. L. Kane, Rev. Modern Phys. 82, 3045 (2010).
\bibitem{Qui}
Xiao-Liang Qi, Shou-Cheng Zhang,  Rev. Moder. Phys. 83, 747 (2011).
\bibitem{wrat2010}
L. Andrew Wray, Su-Yang Xu, Yuqi Xia, Yew San Hor, Dong Qian, Alexei V. Fedorov, Hsin Lin, Arun Bansil, Robert J. Cava and M. Zahid Hasan,  Nat. Phys. 6, 855 (2010).
\bibitem{kriener}
M. Kriener, Kouji Segawa, Zhi Ren, Satoshi Sasaki, and Yoichi Ando,  Phys. Rev, Lett. 106, 127004 (2011).
\bibitem{Hor2010}
Y. S. Hor, A. J. Williams, J. G. Checkelsky, P. Roushan, J. Seo, Q. Xu, H.W. Zandbergen, A. Yazdani, N. P. Ong, and R. J. Cava, {\em Superconductivity in Cu$_x$Bi$_2$Se$_3$ and its Implications for Pairing in the Undoped Topological Insulator} Phys. Rev. Lett. 104, 057001 (2010).
\bibitem{Hor}
 Y.S. Hor, J.G.Checkelsky, D.Qu, N.P.Ong , R.J.Cava, {\em Superconductivity and non-metallicity induced by doping the topological insulators Bi$_2$Se$_3$ and Bi$_2$Te$_3$} Phys. Chem. Solids, 572 - 576  (2011).
\bibitem{hsieh}
Timothy H. Hsieh and Liang Fu, {\em Majorana Fermions and Exotic Surface Andreev Bound States in Topological Superconductors: Application to Cu$_x$Bi$_2$Se$_3$} Phys. Rev. Lett. 108, 107005 (2012).
\bibitem{sat}
Satoshi Sasaki, M. Kriener, Kouji Segawa, Keiji Yada, Yukio Tanaka, Masatoshi Sato, and Yoichi Ando, {\em Topological Superconductivity in Cu$_x$Bi$_2$Se$_3$} Phys. Rev. Lett. 107, 217001 (2011).
\bibitem{par}
Parisa Zareapour, Alex Hayat, Shu Yang F. Zhao, Michael Kreshchuk, Achint Jain, Daniel C. Kwok, Nara Lee, Sang-Wook Cheong, Zhijun Xu, Alina Yang, G.D. Gu, Shuang Jia, Robert J. Cava and Kenneth S. Burch {\em Proximity-induced high-temperature superconductivity in the topological insulators Bi$_2$Se$_3$ and Bi$_2$Te$_3$} Nat Commun 3, 1056 (2012).
\bibitem{koren} G. Koren, T. Kirzhner, E. Lahoud, K. B. Chashka, and A. Kanigel, {\em Proximity-induced superconductivity in topological Bi$_2$Te$_2$Se and Bi$_2$Se$_3$ films: Robust zero-energy bound state possibly due to Majorana fermions} Phys. Rev. B 84, 224521 (2011).
\bibitem{wang}
Mei-Xiao Wang, Canhua Liu, Jin-Peng Xu, Fang Yang, Lin Miao, Meng-Yu Yao, C. L. Gao, Chenyi Shen, Xucun Ma, X. Chen, Zhu-An Xu, Ying Liu, Shou-Cheng Zhang, Dong Qian, Jin-Feng Jia, Qi-Kun Xue, {\em The Coexistence of Superconductivity and Topological Order in the Bi$_2$Se$_3$ Thin Films} Science 336, 52 (2012).
\bibitem{Desheng}
Desheng Kong, Wenhui Dang, Judy J. Cha, Hui Li, Stefan Meister, Hailin Peng, Zhongfan Liu, and Yi Cui, {\em Few-Layer Nanoplates of Bi$_2$Se$_3$ and Bi$_2$Te$_3$ with Highly Tunable Chemical Potential}, Nano Lett. 10, 2245 - 2250 (2010).
\bibitem{tet}
Tetsuroh Shirasawa, Masato Sugiki, Toru Hirahara, Masaki Aitani, Terufusa Shirai, Shuji Hasegawa, and Toshio Takahashi, {\em Structure and transport properties of Cu-doped Bi$_2$Se$_3$ films},  Phys. Rev. B 89, 195311 (2014).
\bibitem{Poole} Charles. P. Poole Jr., Handbook of superconductivity. Academic Press (2000).
\bibitem{daghero} D. Daghero, M. Tortello, G. A. Ummarino and R. S. Gonnelli, Directional point-contact Andreev-reflection spectroscopy of Fe-based superconductors: Fermi surface topology, gap symmetry, and electron–boson interaction, Rep. Prog. Phys. 74 (2011).
\bibitem{schesinger} Z. Schlesinger, R. T. Collins, F. Holtzberg, C. Feild, and S. H. Blanton, Superconducting Energy Gap and Normal-State Conductivity of a Single-Domain YBa$_2$Cu$_3$O$_7$ Crystal, Phys. Rev. Lett. 65, 801 (1990).
\bibitem{dubroka} A. Dubroka, K. W. Kim, M. R$\ddot{o}$ssle, V. K. Malik, A. J. Drew, R. H. Liu, G. Wu, X. H. Chen, and C. Bernhard, Superconducting Energy Gap and c-Axis Plasma Frequency of (Nd,Sm)FeAsO$_0.82$F$_0.18$ Superconductors from Infrared Ellipsometry Phys. Rev. Lett. 101, 097011 (2008).
\bibitem{manzke} R. Manzke, T. Buslaps, R. Claessen and J. Fink, On the Superconducting Energy Gap in Bi$_2$Sr$_2$CaCu$_2$O$_0.8$ Investigated by High-Resolution Angle-Resolved Photoemission, Europhysics lett. 9, 477 (1989).
\bibitem{lee} Mark Lee, D. B. Mitzi, A. Kapitulnik, and M. R. Beasley, Electron tunneling and the energy gap in Bi$_2$Sr$_2$CaCu$_2$O$_x$ Phys. Rev. B 39, 801 (1989).
\bibitem{escudero1} R. Escudero, Rodolfo E. López-Romero, The energy gap of the compound FeSe$_0.5$Te$_0.5$ determined by specific heat and Point Contact Spectroscopy, Solid state Comm. 220,  21–24 (2015).
\bibitem {Fu} Liang Fu and C. L. Kane, Superconducting Proximity Effect and Majorana Fermions at the Surface of a Topological Insulator, Phys. Rev. Lett. 100 (2008).

\bibitem{BTK}G.E. Blonder, M. Tinkham, and T. M. K.lapwijk, Transition from metallic to tunneling regimes in superconducting microconstrictions: Excess current, charge imbalance, and supercurrent conversion, Phycs. Rev. B, 25 (1982).
\bibitem{Zhang2009}
Haijun Zhang, Chao-Xing Liu, Xiao-Liang Qi, Xi Dai, Zhong Fang and Shou-Cheng Zhang, {\em Topological insulators in Bi$_2$Se$_3$, Bi$_2$Te$_3$ and Sb$_2$Te$_3$ with a single Dirac cone on the surface} Nat. Phys. 5, 438 (2009).
\bibitem{Hsieh2009}
D. Hsieh, Y. Xia, D. Qian, L. Wray, J. H. Dil, F. Meier, J. Osterwalder, L. Patthey, J. G. Checkelsky, N. P. Ong, A. V. Fedorov, H. Lin, A. Bansil, D. Grauer, Y. S. Hor, R. J. Cava and  M. Z. Hasan, {\em A tunable topological insulator in the spin helical Dirac transport regime} Nature 460, 1101 (2009).
\bibitem{Wiese1960}
J. R. Wiese and L. Muldawer, {\em Lattice constants of Bi$_2$Se$_3$-Bi$_2$Te$_3$ solid solution alloys} J. Phys. Chem. Solids 15, 13 (1960).
\bibitem{Zhang2009b}
Guanhua Zhang, Huajun Qin, Jing Teng, Jiandong Guo, Qinlin Guo, Xi Dai, Zhong Fang, and Kehui Wua, {\em Quintuple-layer epitaxy of thin films of topological insulator Bi$_2$Se$_3$}, Appl. Phys. Lett. 95, 053114 (2009).
\bibitem{Duming}
Duming Zhang, Jian Wang, Ashley M. DaSilva, Joon Sue Lee, Humberto R. Gutierrez, Moses H. W. Chan, Jainendra Jain, and Nitin Samarth, {\em Superconducting proximity effect and possible evidence for Pearl vortices in a candidate topological insulator} Phys. Rev. B 84, 165120 (2011).
\bibitem{Fan}
Fan Yang, Fanming Qu, Jie Shen, Yue Ding, Jun Chen, Zhongqing Ji, Guangtong Liu, Jie Fan, Changli Yang, Liang Fu, and Li Lu, {\em Proximity-effect-induced superconducting phase in the topological insulator Bi$_2$Se$_3$} Phys. Rev. B 86, 134504 (2012).
\bibitem{Koren2013}
Gad Koren, Tal Kirzhner, Yoav Kalcheim and Oded Millo, {\em Signature of proximity-induced p$_x$ + ip$_y$ triplet pairing in the doped topological insulator Bi$_2$Se$_3$ by the s-wave superconductor NbN}, EPL, 103, 67010 (2013).
\bibitem{yanson}
I.K.Yanson, {\em Nonlinear effects in the electric conductivity of point junctions and electron-phonon interaction in normal metals}, Sov. Phys. JETP 39, 506 (1974).
\bibitem{escudero}
R. Escudero, R. E. Lopez-Romero, and F. Morales, {\em Study of the hidden-order of URu$_2$Si$_2$ by point contact tunnel junctions}, J. Phys.:Condens.Matter 27, 015701 (2015).
\bibitem{fano} U. Fano, {\em Effects of Configuration Interaction on Intensities and Phase Shifts}, Phys. Rev., 124, 1866 (1961).
\bibitem{Parks}
R. D. Parks, {\em Superconductivity}, Marcel Dekker, 1, 597 (1969).
\bibitem{Rauh}
H Rauh, R Geick, H Kohler, N Nucker and N Lehner, {\em Generalized phonon density of states of the layer compounds Bi$_2$Se$_3$, Bi$_2$Te$_3$, Sb$_2$Te$_3$ and Bi$_2$(Te$_0.5$Se$_0.5$)$_3$, (Bi$_0.5$Sb$_0.5$)$_2$Te$_3$} J. Phys. C: Solid State Phys., 14, 2705-2712 (1981).
\bibitem{kirz}
T. Kirzhner, E. Lahoud, K. B. Chaska, Z. Salman, and A. Kanigel, {\em Point-contact spectroscopy of Cu$_0.2$Bi$_2$Se$_3$ single crystals} Phys. Rev. B 86, 064517 (2012).
\bibitem{chen}
X. Chen, C. Huan, Y. S. Hor, C. A. R. S\'{a} de Melo, and Z. Jiang, {\em Point-contact Andreev reflection spectroscopy of candidate topological superconductor Cu$_0.25$Bi$_2$Se$_3$}, arXiv: 1210.6054v1. 22Oct (2012).
\bibitem{niv}
Niv Levy, Tong Zhang, Jeonghoon Ha, Fred Sharifi, A. Alec Talin, Young Kuk, and Joseph A. Stroscio, {\em Experimental Evidence for s-Wave Pairing Symmetry in Superconducting Cu$_x$Bi$_2$Se$_3$ Single Crystals Using a Scanning Tunneling Microscope}, Phys. Rev. Lett. 110, 117001 (2013).
\bibitem{Yamakage2012}
Ai Yamakage, Keiji Yada, Masatoshi Sato, and Yukio Tanaka, {\em Theory of tunneling conductance and surface-state transition in superconducting topological insulators}, Phys. Rev. B 85, 180509 (2012).

\end{thebibliography}
\end{document}